\documentclass[%
reprint,
amsmath,amssymb,
aps,
prstab,
]{revtex4-1}

\usepackage{graphicx}
\usepackage{dcolumn}
\usepackage{bm}
\usepackage{siunitx}
\usepackage{wasysym}
\usepackage{color}
\usepackage{hyperref}

\begin{document}


\title{Characterization of a plasma window as a membrane free transition 
	\\between vacuum and high pressure}

\author{B. F. Bohlender}
 \email{bohlender@iap.uni-frankfurt.de}

\author{A. Michel}%
\author{J. Jacoby}
 \altaffiliation[Also at ]{GSI Helmholtzzentrum für Schwerionenforschung}
\author{M. Iberler}
\affiliation{%
 IAP, Institute for Applied Physics\\
 Goethe Universität Frankfurt}%
\author{O. Kester}
\affiliation{
 TRIUMF \\
 Vancouver B.C., Canada
}

\date{\today}

\begin{abstract}
A plasma window (PW) is a device for separating two areas of different pressures while
letting particle beams pass with little to no loss.
It has been introduced by A. Hershcovitch \cite{HER95}.
In the course of this publication, the properties of a PW with apertures of
\SI{3.3}{\milli\meter} and \SI{5.0}{\milli\meter} are presented.
Especially the link between the pressure properties relevant for applications in
accelerator systems and the underlying plasma properties depending on external
parameters are presented.

As working gas, a \SI{98}{\percent}{Ar}-\SI{2}{\percent}H$_2$ mixture has been used due
to the intense Stark broadening of the H$_\beta$-line and the well-described {Ar}
characteristics, enabling an accurate electron density and temperature analysis.
At the low pressure side around some \SI{}{\milli\bar}, high-pressure values
reached up to \SI{750}{\milli\bar} while operating with volume flows between
\SI{1}{slm} and \SI{4}{slm} (\textit{standard liter per minute}) and discharge
currents ranging
from \SI{45}{\ampere} to \SI{60}{\ampere}.
The achieved ratios between high and low pressure with an active discharge range from
40 to 150.
This is an improvement of a factor up to 12 over the performance of an ordinary
differential pumping stage of the same geometry.

Unique features of the presented PW include simultaneous plasma parameter
determination and the absence of ceramic insulators between the cooling plates.
Optical analysis reveals no significant damage or wear to the components after an operation
time well over \SI{10}{\hour}, whereas the cathode needle needs replacement after
\SI{5}{\hour}.

\end{abstract}

\maketitle

\section{Introduction\label{sec:level1}}
Modern atomic and particle physics call for intense and brilliant particle beams with
high energies, while specific applications require the separation between the
accelerator's vacuum and areas of higher pressures, e.g. gas strippers or
experimental chambers.
Especially when producing radioactive isotopes in the course of an experiment,
the accelerator needs to be shielded from debris to prevent contamination.

Technological applications aside from the usage in accelerator systems include atmospheric
electron welding, cutting and surface modification also calling for a reliable
vacuum-atmosphere interface.
Additionally, x-ray microscopy in living cells suffers from the degrading of the used
SiN-Windows and their small size, again demanding a long living interface.

Conventional means of beam transfer from vacuum to regions of high pressure is usually
archived by using metallic membranes or differential pumping stages.
Growing beam intensities limit the usage of membrane windows due to their destruction
by beam interaction, while differential pumping stages grow unacceptable long for high
pressures.
Therefore, the physical community seeks for material free separation of high pressure
regions in accelerators.

A possible alternative has been proposed by A. Hershcovitch in 1995 \cite{HER95}: 
The improvement of a differential pumping stage by introducing an arc discharge
into the stage. 
This so called plasma window (PW) uses a cascaded arc discharge \cite{MAE56} for 
the connection between vacuum and high pressure.
The plasma is ignited in a channel connecting the different pressure areas formed by
insulated copper discs.
These discs stabilize the arc due to spatial and thermal confinement and need to be
water cooled due to the high power dissipation,
$P_\mathrm{loss}\leqslant \SI{1}{\kilo\watt\per\centi\meter}$, and temperature
$T\simeq\SI{1e4}{\kelvin}$ of the discharge \cite{MAE56}.

While the pressure ratio characteristics of the PW have been subject to 
research and simulations for apertures between \SI{2}{mm} and \SI{8}{mm} 
\cite{NAM16, SHI14, HER95, HUA14, KUB13, KRA07}, the underlying plasma 
properties have only sparsely been published, one exception being \cite{VIJ10}
for the case of a hydrogen discharge.
For future applications and development of the plasma window technology, insight
into the linking of external parameters to the inner plasma characteristics and the
achievable pressure differences needs to be researched.

Transmission of particles through a PW has been shown for electrons, VUV and soft
x-rays, although these publications \cite{HER95, HER98, PIN01, HER05, GIU11}
only state particles or photons were transmitted through the PW.
Studies on the transmission characteristics or the influence of the plasma
on particle beam properties like average charge state or emittance have not been
conducted and/or published yet.

This paper presents the basic gas dynamic and plasma properties inside the PW
with apertures of \SI{3.3}{\milli\meter} and \SI{5.0}{\milli\meter},
the plasma properties and their influence on the achievable pressure ratios of the PW.

Of particular interest for further applications of the plasma window technology is
the improvement of the pressure ratio in relation to a differential pumping stage.
The improvement in sealing of the plasma window compared to a differential pumping stage
can be expressed as
\begin{equation}
	q_n = \frac{p_H}{p_{H,0}} \label{eq:qn}
\end{equation}
where $p_H$ is the high side pressure with active discharge and $p_{H,0}$ is that of
a conventional pumping stage with the same geometric properties as the PW at
fixed particle flow.
$q_n$ will later on be used to classify the performance of the presented PW.


\subsection{Plasma physics \label{sec:plasma}}
As the PW's sealing mechanism is supposed to originate in the heating of the working
gas \cite{HER95}, a look into the thermodynamic properties of plasmas is
worthwhile.

A plasma is composed of different particle species, usually electrons, ions and neutral
atoms or molecules.
Generally speaking, the electrons usually carry significant higher energies than the
heavier species.
If all species carry locally the same kinetic energy, the plasma is in a state called
\textit{local thermodynamic equilibrium} (LTE).

For a LTE to prevail, the electron density $n_e$ needs to be high enough to ensure
sufficient energy transfer from the fast and light electrons to the heavier species.
\cite{SAL98} formulated an expression to calculate the necessary electron
density:
\begin{equation}
	\frac{n_\mathrm{crit}}{\mathrm{cm}^{-3}} = 
		1\times 10^{14} \left(\frac{k_BT_e}{\mathrm{eV}}\right)^3
		\left(\frac{\epsilon_\mathrm{ion}}{k_BT_e}\right)^{5/2}
	\label{eq:necrit}
\end{equation}
Where $k_BT_e$ is the mean kinetic energy of the electrons, $\epsilon_\mathrm{ion}$ is
the ionization energy of the species under question.

For an Argon plasma with electron temperatures around
$k_BT_e\approx \SI{1}{\electronvolt}$, this yields a critical electron density around
$n_\mathrm{crit}=\SI{9.9e16}{\per\cubic\centi\meter}$.

It's worth to stress that only if the observed electron density is higher than
$n_\mathrm{crit}$, the heavy particle temperature is equivalent that of the electrons.
Otherwise, no accurate determination of the heavy particle temperature
is possible, but it increases with increasing electron density.

\section{Experimental setup and data evaluation\label{sec:window}}

\subsection{Experimental setup}
\begin{figure*}[h!tb]
	\centering
	\includegraphics[width=17.2cm]{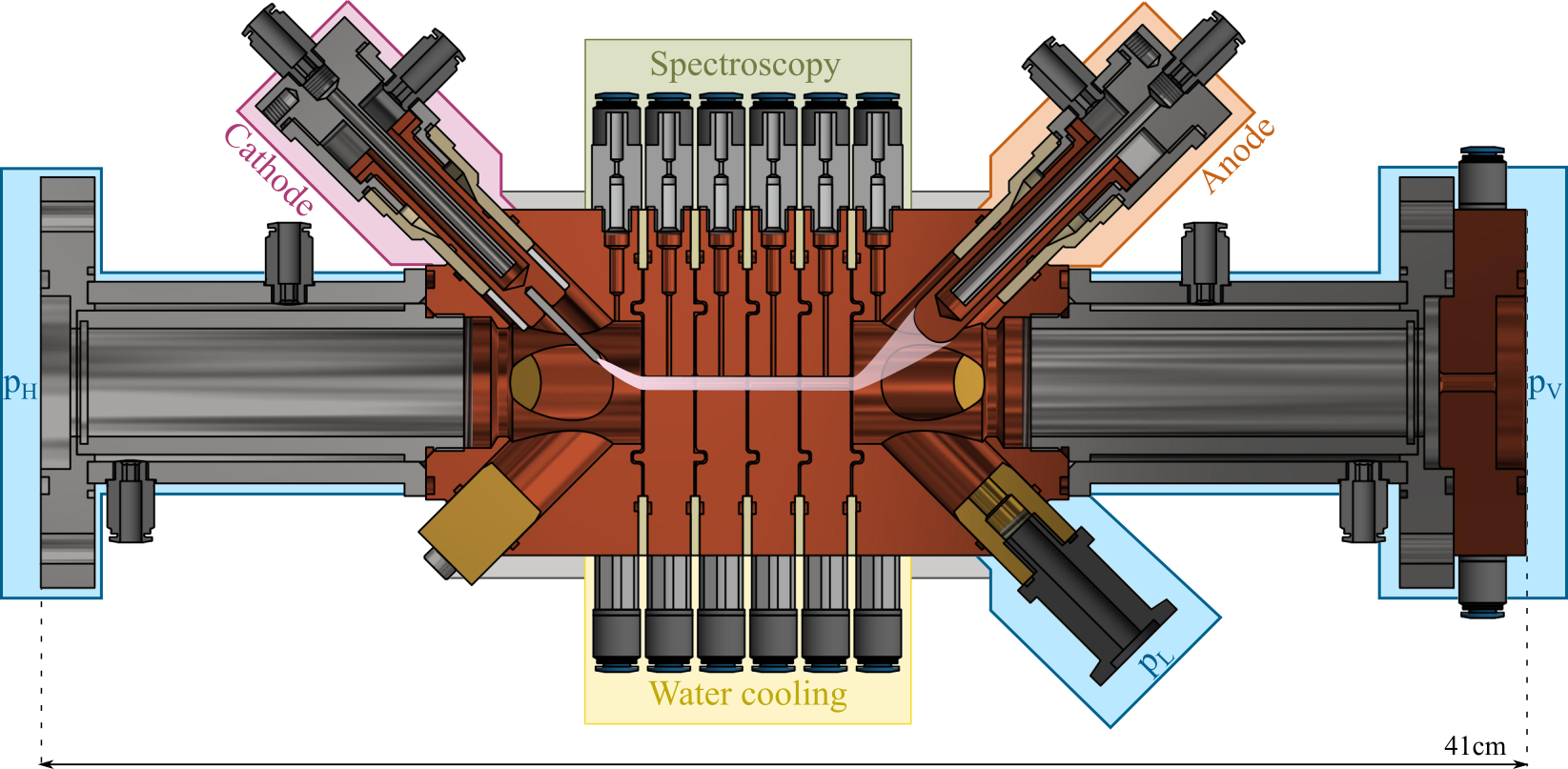}
	\caption{\label{fig:QS}Cross section of the PW in use, the geometrical 
		parameters are given in table \ref{tab:geo}.
		The high pressure side $p_H$ is located to the left.
		The active discharge is indicated as a bright pink area.}
\end{figure*}
Figure$\,$\ref{fig:QS} shows the cross section of the PW developed at IAP.
On the left side of the schematic, the recipient containing the high pressure
$p_H > \SI{100}{mbar}$ is shown.
The PW itself consists of the cathode section upstream, the anode section
downstream of the plasma and four copper plates connecting cathode and anode section.
The cathode, formed from a $W-La_2O_3$ pin, is being held by a water cooled copper body
which is insulated by a PEEK-sleeve and MACOR hood from the body of the setup.
The anode is located at the lower pressure side and made from copper.

The cooling plates are made from copper and separated by PEEK-Spacers. 
These spacers usually suffer damage from the high intensity optical radiation 
emitted by the plasma and need to be shielded from this radiation.
Designs by other groups achieve this by using boron nitride washers, while the here
presented PW implements a tongue-and-groove design, which is less prone to 
failure and easier to manufacture than brittle ceramics.

For operation of the PW, a gas mixture of \SI{98}{\percent} Ar + \SI{2}{\percent} H$_2$
is fed into the setup at several flow rates in the range from \SI{1}{slm} up to
\SI{4}{slm}.
This is done to enable an accurate temperature determination via \textit{Boltzmannplot}
techique using up to 11 ArII-Lines, while the H$_2$-addition allows for precise electron
density calculation.
In order to record the necessary spectra, each cooling plate features an optical port.
This allows the simultaneous acquisition of spectral data emerging from different points
along the discharge axis at the same time.
Combined, these measurements allow for insight into the plasma characteristics under
different global parameter settings, such as current, pressure and volume flow rate.

\begin{table}[h!tb]
	\centering
	\caption{\label{tab:geo} Geometry and setup data for the IAP PW}
	\begin{tabular}{l|r}
		\toprule
		Length over all & \SI{41}{\centi\meter}\\
		Channel length & \SI{59}{\milli\meter}\\
		Channel diameter & 3.3 and \SI{5.0}{\milli\meter}\\
		Current & 40 \ldots\SI{60}{\ampere}$_=$\\
		Voltage & $\leq\,$\SI{150}{\volt}$_=$ \\
		Gas mixture & Ar + 2\% H$_2$\\
		Pumping speed & $\approx \SI{400}{slm}$\\
		\toprule
	\end{tabular}
\end{table}
Table \ref{tab:geo} summarizes the experimental parameters under which 
the presented measurements have been conducted.
The data presented in this contribution was collected by using a scroll pump at 
the low pressure side.

\subsection{Electron density and temperature evaluation\label{sec:set_plasm}}
Plasma characterization is done by spectroscopic methods.
For the calculation of the electron density $n_e$ from the H$_\beta$-broadening, the
following formula from \cite{GIG03} is used:
\begin{equation}
\mathrm{FWHM} = 2\gamma = 
\SI{4.8}{\nano\meter}\left(\frac{n_e}{\SI{1e17}{\per\cubic\centi\meter}}\right)^{0.68116}
\label{eq:hbeta}
\end{equation}
In Eq.$\,$\ref{eq:hbeta} FWHM is the full width at half maximum of the broadened line profile.
The accuracy of the measured electron density is typically better than \SI{10}{\percent}.

As for the electron temperature determination, the \textit{Boltzmannplot} \cite{KUN09}
method with selected ArII-lines is used.
The used lines and some of their relevant quantities are given in Tab. \ref{tab:linien}.
In order to achieve a good temperature estimation, spectra with three different
wavelength frames were recorded.
The optical setup's response was adjusted for its sensitivity in the full range of the observed
lines.
By doing so, up to 11 ArII-lines could be used for the electron temperature
determination at any given set of parameters and optical ports, resulting in an
accuracy around \SI{7.5}{\percent}.
\begin{table}[h!tb]
	\centering
	\caption{ArII-lines used for the temperature determination via Boltzmann plot
		technique, taken from \cite{NIST_ASD}.
		\label{tab:linien}}
	\begin{tabular}{l|r|r|r|c|c}
		$\lambda$/nm & $A$/s$^{-1}$ & $g_p$ & $E_p$/eV & Configuration & Term\\
		\toprule
		457.93 & 8.0e+7 & 2 & 19.973& $3s^23p^4(^3P)4p$& $^2S^\circ_{1/2} $\\
		458.98 & 6.64e+7 & 6 & 21.127& $3s^23p^4(^1D)4p$& $^2F^\circ_{5/2} $\\
		460.96 & 7.89e+7 & 8 & 21.143& $3s^23p^4(^1D)4p$& $^2F^\circ_{7/2} $\\
		465.79 & 8.92e+7 & 2 & 19.801& $3s^23p^4(^3P)4p$& $^2P^\circ_{1/2} $\\
		480.60 & 7.80e+7 & 6 & 19.223& $3s^23p^4(^3P)4p$ &$^4P^\circ_{5/2} $\\
		484.78 & 8.49e+7 & 2 & 19.305& $3s^23p^4(^3P)4p$ &$^4P^\circ_{1/2} $\\
		487.98 & 8.23e+7 & 6 & 19.680& $3s^23p^4(^3P)4p$ &$^2D^\circ_{5/2} $\\
		496.50 & 3.94e+7 & 4 & 19.762& $3s^23p^4(^3P)4p$ &$^2D^\circ_{3/2} $\\
		497.21 & 9.7e+6 & 2 & 19.305& $3s^23p^4(^3P)4p$& $^4P^\circ_{1/2} $\\
		500.93 & 1.51e+7 & 6 & 19.223& $3s^23p^4(^3P)4p$& $^4P^\circ_{5/2} $\\
		501.71 & 2.07e+7 & 6 & 21.127& $3s^23p^4(^1D)4p$ &$^2F^\circ_{5/2} $\\
		\toprule
	\end{tabular}
\end{table}

The used spectra were recorded through radial windows in the PW,
see Fig.$\,$\ref{fig:QS}, and a \SI{0.5}{\meter}-Monochromator.

\subsection{Acquisition of electrical and pressure data\label{sec:set_pres}}

The discharge current and voltage was recorded from the main power supply.
The power supply has an accuracy of $\Delta U = \SI{1.25}{\volt}$ and 
$\Delta I = \SI{0.3}{\ampere}$.

Pressure values at the cathode and anode, see $p_H$ and $p_L$ in Fig.$\,$\ref{fig:QS},
were taken with two \textit{Agilent PCG-750} manometers.
The pressure in the pumping recipient, $p_V$ in Fig.$\,$\ref{fig:QS}, was recorded using a
\textit{Pfeiffer PKR 251} manometer.
The used manometers have typical errors of \SI{5}{\percent} (Agilent) and \SI{30}{\percent}
(Pfeiffer) respectively.

\section{Results and discussion\label{sec:ResDis}}

\subsection{Electrical parameters \label{sec:res_elec}}
\begin{figure}[h!tb]
	\centering
	\includegraphics[width=0.48\textwidth]{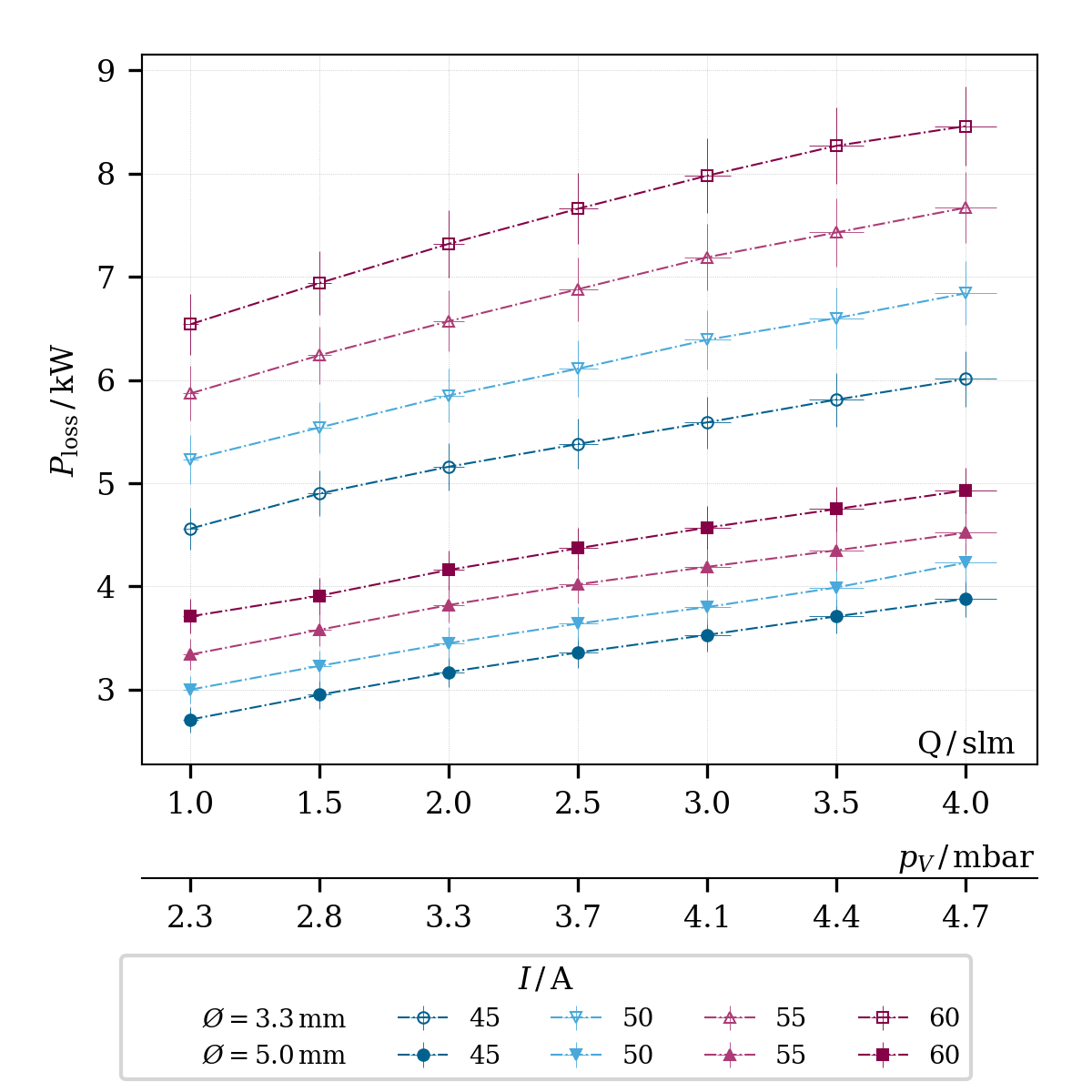}
	\caption{Electrical power $P_\mathrm{loss}$ in dependence of
		current and pressure.
		Errorbars in x-direction refer to the error in $Q$.}
	\label{fig:P_loss}
\end{figure}

The electrical measurements show that the power needed for sustaining the discharge
scales with pressure and current but drops with increasing aperture.
The dependence is shown in Fig.$\,$\ref{fig:P_loss}.

The increase of $P_\mathrm{loss}$ with increasing current and pressure and decreasing
aperture originates from the increase of neutral particles within the channel,
see Sec.$\,$\ref{sec:pressure} and \cite{FRI11}.
As a consequence, the plasma's electrical resistance grows \cite{BOU94},
therefore a higher voltage is necessary to sustain the discharge the same current.

\subsection{Plasma parameters\label{sec:res_plasma}}
\begin{figure}[h!tb]
	\centering
	\includegraphics[width=0.45\textwidth]{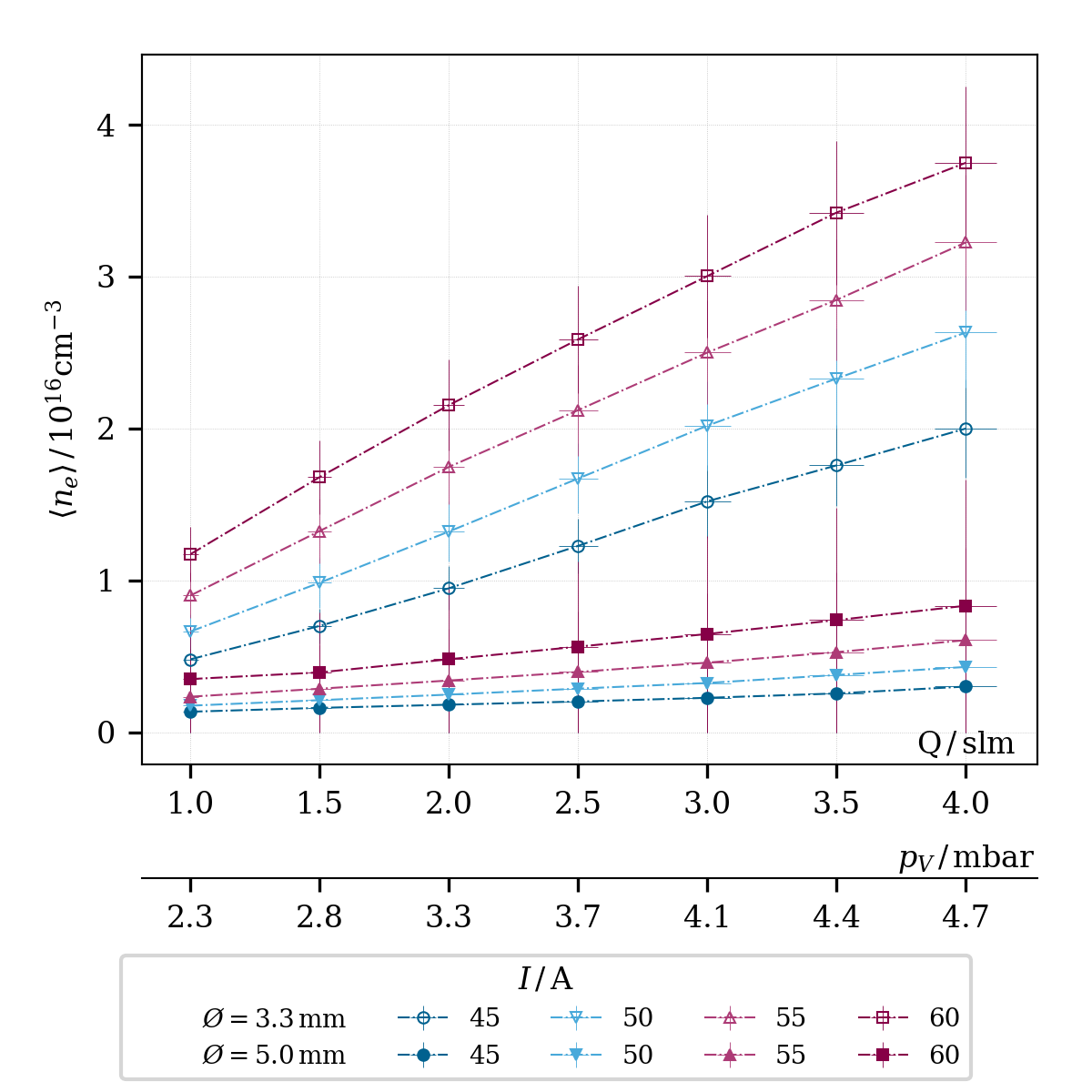}
	\caption{Electron density $\left<n_e\right>$ averaged along discharge channel 
		in dependence of current and pressure.
		Errorbars in x-direction refer to the error in $Q$.}
	\label{fig:ne_avg}
\end{figure}

\begin{figure}[h!tb]
	\centering
	\includegraphics[width=0.45\textwidth]{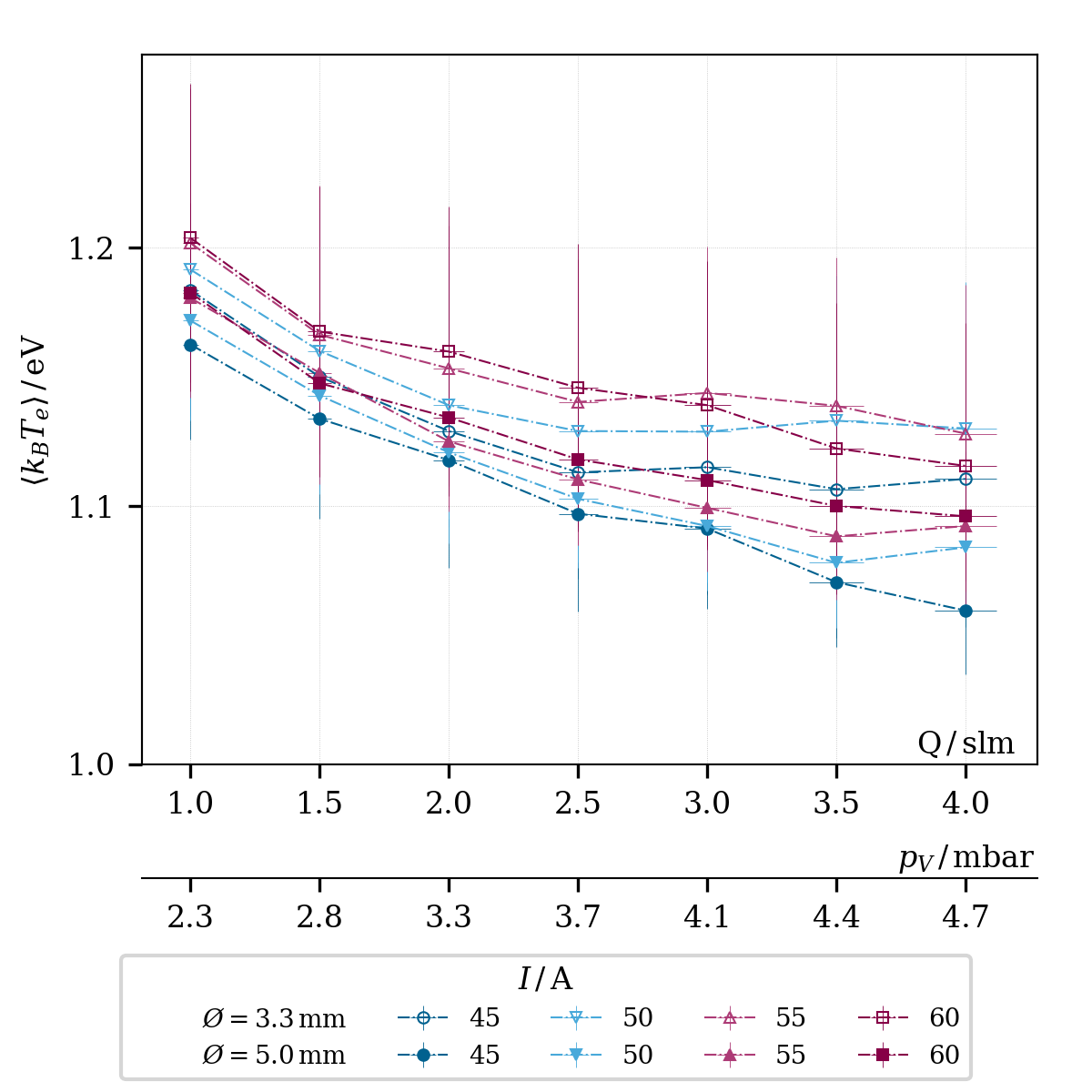}
	\caption{Electron temperature $\left<k_BT_e\right>$ averaged along discharge channel
		in dependence of current and pressure.
		Errorbars in x-direction refer to the error in $Q$.}
	\label{fig:Te_avg}
\end{figure}

The recorded spectra are used to calculate the electron density $n_e$ and temperature
$k_BT_e$ at every observation point.
Calculations are done according to the description given in Sec.\ref{sec:set_plasm}.
These calculated values are averaged over the discharge axis and shown in
Fig.'s$\,$\ref{fig:ne_avg} and \ref{fig:Te_avg} as $\left<n_e\right>$ and
$\left< k_BT_e\right>$ to illustrate the collective behaviour of these quantities.

Electron density $\left<n_e\right>$ scales with current and volume flow, which is in
good agreement with \cite{RAI91} and \cite{KRO88_II}.
The increase of $\left<n_e\right>$ with the volume flow originates from the higher
number of particles inside the plasma and indicates a constant degree of ionization.
Due to the limited discharge cross section, an increase of the current gives raise to
a higher current density, thus a higher number of electrons inside the discharge.

The maximal density is \SI{3.75e16}{\per\cubic\centi\meter}, which is not sufficient
for the plasma to be in a LTE, see Sec.$\,$\ref{sec:plasma}.
Therefore, no valid statement can be made about the heavy particle tempearture, but
according to \cite{FRI11} and \cite{RAI91}, the heavy particle temperature
increases with increasing electron denisty.

The electron temperature $\left< k_BT_e\right>$ is decreasing with increasing particle flow.
With more particles inside the plasma volume, the collision frequency increases, thus
electrons transfer more energy to heavy particles, reducing the electron temperature
\cite{RAI91}.
On the contrary, an increase of the discharge current seems to have no significant effect
on the electron temperature.

\subsection{Lifetime\label{sec:Res_life}}
Optical investigation of the cooling plates and the anode reveals no significant damage
after well over \SI{10}{\hour} of operation.

\begin{figure}[h!tb]
	\centering
	\includegraphics[width=0.45\textwidth]{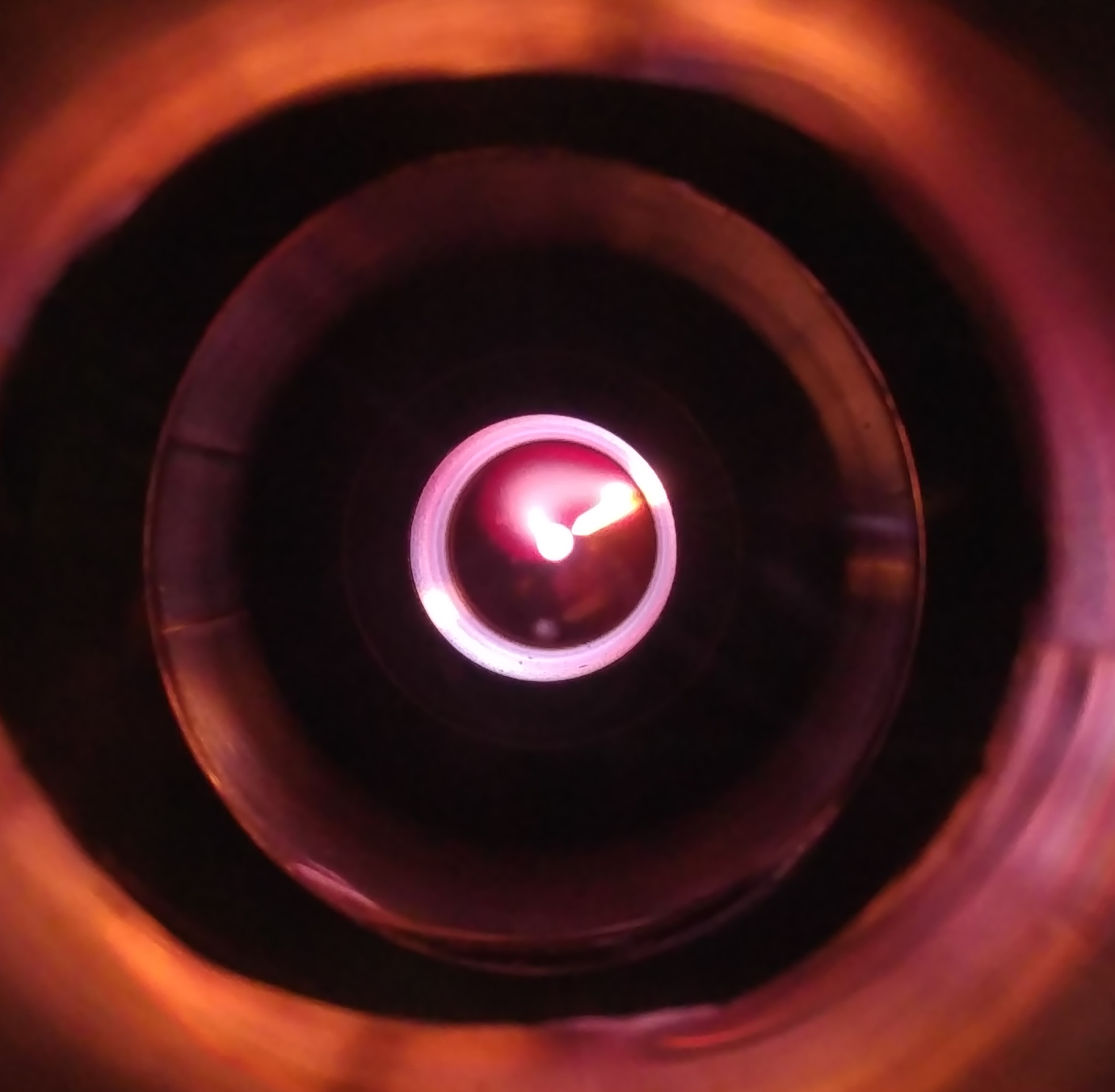}
	\caption{View onto the discharge from the cathode side.
		The plasma filled discharge channel is the bright central circle, the cathode
		sticks out from the upper right side.
		Visible is the anchor point which has already moved up the tip.}
	\label{fig:abbrand}
\end{figure}

In contrast to this stability, the discharge's anchor point at the cathode tip 
drifts towards the cathode body, as indicated in Fig.$\,$\ref{fig:abbrand}.
The drift begins to become visible after about \SI{1}{\hour} of operation.
Since the cathode's body is made from copper, the tip needs to be replaced before the
discharge reaches the body, which would inevitably melt otherwise.
Due to the needed replacement, the operation time is limited to \SI{5}{\hour} at a
stretch.

Our assumption about the cause of this drift is based on the tip's material.
Due to the high temperature of the needle under operation, the Lanthanoxide may burn
out from the tip, leaving pure tungsten material.
Since Tungsten has a higher work function than the Tungsten-Lanthanoxide compound, the
anchor drifts away from the burnt out areas.

A possible enhancement of the lifetime might thus arise from using pure Tungsten tips
in the future, which comes with the drawback of the then needed increased electrical
power to sustain the discharge.

\subsection{Pressure parameters\label{sec:pressure}}

Considering the particle flow $\Gamma = Q\times n$ and using Hagen-Poiseuille's
law for compressible fluids as an approximation for the gas flow throughout the Window,
one can derive
\begin{equation}
	\Gamma = \frac{\pi R^4}{32 l}\left(\frac{p_H^2-p_V^2}{p_V}\right) \frac{n}{\eta}
	\label{eq:Gamma}
\end{equation}
where $l$ is the length of the channel, $R$ its radius, $n$ the paritcle denisty and
$\eta$ the viscosity of the fluid under consideration.

Rearranging Eq.$\,$\ref{eq:Gamma} under the assumption that $p_H^2 \gg p_V^2$, an
expression for $p_H$ can be formulated:
\begin{equation}
p_H \approx \frac{1}{\pi R^2} \sqrt{32 \,l \,\Gamma\pi\, p_V}\,\sqrt{\frac{\eta}{n}}
\label{eq:ph}
\end{equation}
Since the viscosity of a gas increases with its temperature while its denisty decreases
\cite{BOU94}, an increase of the gas temperature induces higher value for $p_H$, if the
particle flow and thus $p_V$ are kept constant.

\begin{figure}[h!tb]
	\centering
	\includegraphics[width=0.45\textwidth]{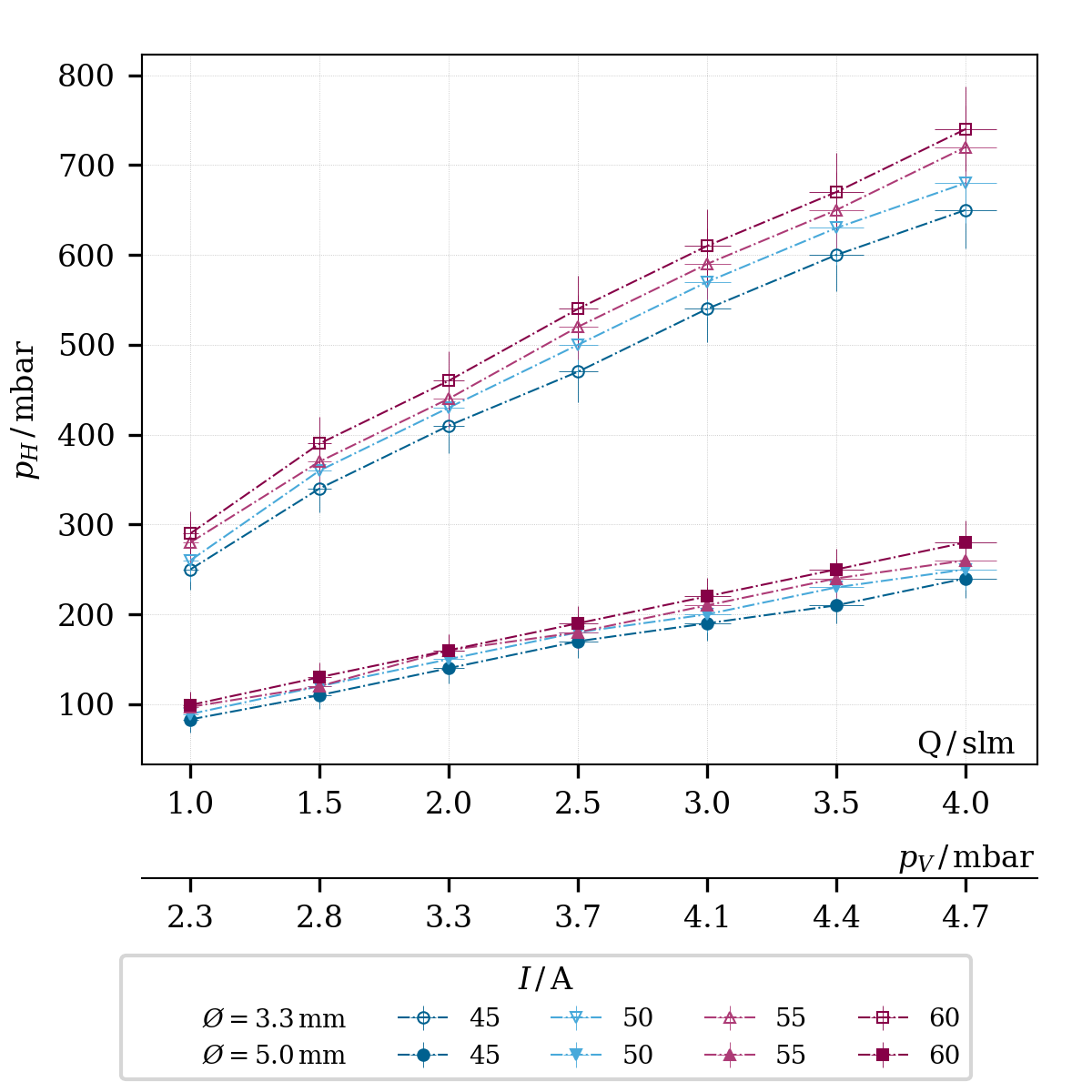}
	\caption{High pressure $p_H$ in dependence of the discharge current $I$
		and volume flow $Q$.
		Errorbars in x-direction refer to the error in $Q$, the systematic error in
		$p_V$ is \SI{30}{\percent}.}
	\label{fig:phpv}
\end{figure}

The scaling of high side pressure $p_H$ with applied volume flow $Q_0$ and discharge
current $I$ is depicted in Fig.$\,$\ref{fig:phpv}. 
Clearly the channel's aperture has the largest influence on the obtained high
pressure $p_H$.
The shown weak scaling of $p_H$ with $I$ is in good agreement with the data published
in \cite{NAM16} for currents above \SI{10}{\ampere}.

As the viscosity of gases increases with increasing temperature at most by a factor 3,
the viscosity's influence on the performance of the PW is limited.

\begin{figure}[h!tb]
	\centering
	\includegraphics[width=0.45\textwidth]{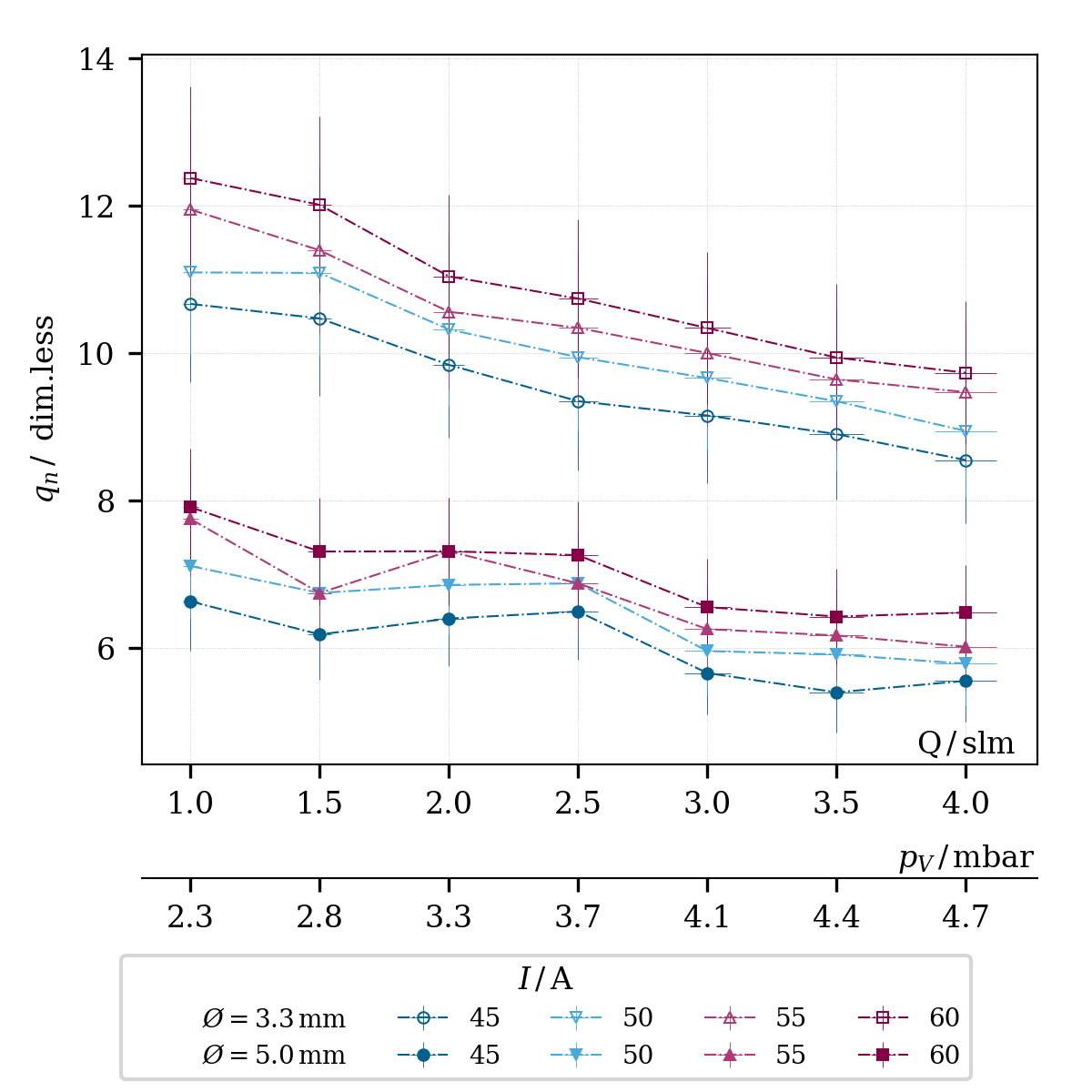}
	\caption{Normalized pressure quotient $q_n=\frac{q}{q_0}$ in dependence of
		discharge current $I$ and volume flow $Q$.
		Errorbars in x-direction refer to the error in $Q$.}
	\label{fig:qnpv}
\end{figure}

The ratio $q_n=\frac{p_H}{p_{H,0}}$ characterizes the sealing improvement of the PW
over the same setup operated as an ordinary differential pumping stage.
This quantity could be used in future publications for comparing different PW from
varying groups in terms of their performance.
Its behaviour with external parameters is depicted in Fig.$\,$\ref{fig:qnpv}.

An improvement of a factor up to 12 can be seen from the data presented in
Fig.$\,$\ref{fig:qnpv}.
As with high pressure, the smaller channel diameter yields a better performance.
For both used diameters $q_n$ scales with the current and against the volume flow.

\section{Conclusion}
It has been shown that the PW developed at IAP is capable of maintaining a pressure
difference up to $p_H={}$\SI{750}{mbar} to $p_V={}$\SI{4.7}{mbar} for over \SI{5}{h}
at continuous operation with little to no signs of erosion.
This is achieved using a single scroll pump with a pumping speed of \SI{200}{slm}.
The sealing is caused by increasing the heavy particle temperature so that the particle
density inside the channel decreases and the viscosity of the gas increases.
Heating of the particles is achieved by increasing the electron density, which causes a
more efficient energy transfer from the electrons to the heavy particles.

The water cooling of the downstream recipient proved crucial, as the hot gas streaming
out of the discharge heats the recipient up to a point where the sealing starts to
fail.
For prolonged operation of the window, other cathode materials than $W-La_2O_3$ need to
be tested since this compound proved not to be stable enough of long term operation.

In order to characterize the PW's performance, additional tests with different working
gases are scheduled for the near future.
Additionally, the test of pure tungsten cathode pins and studies of the influence
of the number of cathodes on the PW's lifetime are on their way.
Furthermore, an upgrade of the vacuum system will be performed, presumably increasing the
achievable pressure difference.

For the future, beam interaction tests are planned to specify the transmission
properties of the PW for different particles.
%

%
%

\section*{Acknowledgments}
The here presented work would not have been possible without the support of
the BMBF, Ref.no 05P15 RFRBA, HIC for FAIR and HGS-Hire.

\bibliography{PRAB-BIB}
\end{document}